# Accelerated Discovery of Efficient Solar-cell Materials using Quantum and Machine-learning Methods

Kamal Choudhary[1], Marnik Bercx[2], Jie Jiang[3], Ruth Pachter[3], Dirk Lamoen[2], and Francesca Tavazza[1]

1 Materials Science and Engineering Division, National Institute of Standards and Technology, Gaithersburg, Maryland 20899, USA.

2 EMAT, Department of Physics, University of Antwerp, Groenenborgerlaan 171, 2020 Antwerp, Belgium.

3 Materials Directorate, Air Force Research Laboratory, Wright–Patterson Air Force Base, Ohio 45433, USA.

## Abstract:

Solar-energy plays an important role in solving serious environmental problems and meeting high-energy demand. However, the lack of suitable materials hinders further progress of this technology. Here, we present the largest inorganic solar-cell material search to date using density functional theory (DFT) and machine-learning approaches. We calculated the spectroscopic limited maximum efficiency (SLME) using Tran-Blaha modified Becke-Johnson potential for 5097 non-metallic materials and identified 1997 candidates with an SLME higher than 10%, including 934 candidates with suitable convex-hull stability and effective carrier mass. Screening for 2D-layered cases, we found 58 potential materials and performed $G_0W_0$ calculations on a subset to estimate the prediction-uncertainty. As the above DFT methods are still computationally expensive, we developed a high accuracy machine learning model to pre-screen efficient materials and applied it to over a million materials. Our results provide a general framework and universal strategy for the design of high-efficiency solar cell materials. The data and tools are publicly distributed at: https://www.ctcms.nist.gov/~knc6/JVASP.html, https://www.ctcms.nist.gov/jarvisml/, https://jarvis.nist.gov/ and https://github.com/usnistgov/jarvis .

# Introduction

Solar-cells[1, 2] are one of the most promising sustainable energy alternatives. Their success, however, is heavily dependent on finding suitable materials. Despite substantial progress in identifying solar-cell materials, the field is facing a formidable challenge due to a lack of high-quality and large-volume frequency-dependent dielectric function data. Recently, systematic investigations for photovoltaic (PV) materials have gained increasing interest in the density functional theory (DFT) community[3], leading to the identification of candidate materials like several chalcopyrites[4, 5], tetrahedrite[6], Cu-In halide perovskites[7] and layered perovskites (Ruddlesden–Popper and Dion–Jacobson phases) [8]. Most of these materials have been predicted to be suitable for photovoltaics using the spectroscopic limited maximum efficiency (SLME) approach[4]. However, the number of known inorganic materials (such as those in the ICSD database[9]) is on the order of hundreds of thousands, whereas the frequency-dependent dielectric function required for the SLME is only reported for a couple hundred, computed for example at the computationally intensive $G_0W_0$-BSE level of theory. In other words, a large, systematic 'database' of potential efficient materials is still missing and highly desirable. Such a dataset is the first step towards the development of any data-analytics or machine learning model as well[10].

Many-body perturbation theory approaches (such as GW[11] and GW-BSE[12]) are generally considered to be necessary to obtain accurate efficiencies because they accurately predict band gaps and frequency-dependent dielectric functions. However, meta-GGA based methods, such as the Tran-Blaha modified Becke-Johnson (TBmBJ) potential[13], have been recently shown to achieve comparable accuracy in evaluating the same quantities at a significantly reduced computational cost, enabling the calculation of the frequency-dependent dielectric function and bandgap for thousands of inorganic crystalline materials[14]. The next step is to investigate if this data can be used to identify novel high solar-efficiency[15] materials. One of the earliest selection-metrics for identifying solar cell materials was introduced by Shockley-Queisser (SQ)[15], which utilized information about the bandgap, blackbody radiation, and solar spectrum to estimate an upper limit for the efficiency. However, the SQ formalism did not consider the absorption coefficient and thickness of the absorber material. Yu et al.[4] introduced the spectroscopic limited maximum efficiency (SLME) approach, applied it to 260 materials and predicted 20 high-SLME materials. The SLME overcame the shortcomings of the SQ limit by incorporating the absorptivity and therefore essentially taking dipole matrix elements and thickness into account. Additional investigations are needed to examine various other factors that may impact the efficiency, such as effective carrier mass and lifetime of charge carriers[16], internal efficiency of the cell[17], cost of the materials, defect-tolerance[18], thermal degradation tolerance, and chemical inertness, which are also critically important aspects when designing a photovoltaic device. While meta-GGA methods allow the investigation of hundreds to thousands of materials, the computational cost is still too high for tackling $10^{100}$ possible materials[19]. Recently, machine learning for materials modeling has emerged as a promising new solution to this problem. Having a systematic dataset like JARVIS-DFT enables the application of machine learning techniques. The JARVIS-DFT database contains about 30000 bulk and 800 low-dimensional materials with their DFT-computed structural, energetics[20], elastic[21], optoelectronic[14] , thermoelectric[22] and topological material[23] properties.

In this work, we introduce a workflow for identifying potential solar absorber materials by performing a high-throughput DFT screening based on the SLME, effective mass of charge carriers and the convex-hull stability for non-metallic systems and machine learning. The use of such a workflow allows us to narrow down the list of materials to a manageable number so that it is realistic to perform experimental investigations of their solar cell efficiencies. High-throughput DFT based screening has been successfully used by several researchers to screen high-performance materials such as in AFLOW[24], Materials-Project (MP)[25] and Open Quantum Materials Database (OQMD)[26]. But, due to the inherent issue of bandgap underestimation in conventional DFT methods, metrics such as SLME cannot be accurately predicted and hence not available in the above databases. In the Harvard Clean Energy Project (HCEP)[27], Aspuru-Guzik and co-workers used similar high-throughput screening approaches to identify high-efficiency molecular materials. Out of 30000 solid-state materials available in JARVIS-DFT, we have Tran-Blaha modified Becke-Johnson potential (TBmBJ)[13] data for 12881 materials only. Ignoring metallic systems, we calculated the TBmBJ SLME for 5097 materials. Out of these materials, 1997 candidates have an SLME above the threshold of 10%. We further narrowed the search by screening for effective carrier mass less than 1.0 $m_0$ , where $m_0$ is the mass of a free electron, and convex hull stability (<0.1 eV/atom), leading to 934 candidates. Our screening methodology is then applied to the search for solar-cell materials with 2D character, as this could combine the technological applicability of both classes of materials. We found 58 such materials and performed $G_0W_0$ with and without spin-orbit coupling (SOC) calculations on a subset of them to evaluate the uncertainty related to the neglection of many-body effects. Lastly, we developed a high accuracy machine learning tool based on the classification method to pre-screen materials in terms of high-SLME and we applied it to over a million materials available through large crystallographic and DFT databases such as AFLOW[24], Materials-project[25], Open Quantum Materials Database (OQMD)[26], Crystallography Open Database (COD)[28] and JARVIS-DFT. We made all the predicted materials publicly available through the website: https://jarvis.nist.gov/.

## Computational methodology

All DFT calculations were carried out with Vienna *ab initio* simulation package (VASP)[29, 30] using projected augmented wave (PAW) formalism. Please note commercial software is identified to specify procedures. Such identification does not imply recommendation by the National Institute of Standards and Technology. The k-point and plane-wave cut-off convergence for each material are obtained using the workflow detailed in Ref.[31] We included three times as many empty conduction bands as valence bands, which is necessary for calculating the electronic transitions over an adequate energy range. We choose 5000 energy grid points to have a sufficiently high resolution in dielectric function spectra. The imaginary part is calculated as:

$$\varepsilon_{\alpha\beta}^{(2)}(E) = \frac{4\pi^2 e^2}{\Omega^2} \lim_{q\to 0} \frac{1}{q^2} \sum_{c,v,\vec{k}} 2w_{\vec{k}} \delta\left(\xi_{c\vec{k}} - \xi_{v\vec{k}} - E\right) \langle \Psi_{c\vec{k}+\vec{e}_\alpha q} | \Psi_{v\vec{k}} \rangle \langle \Psi_{v\vec{k}} | \Psi_{c\vec{k}+\vec{e}_\beta q} \rangle^* \qquad (1)$$

where $e$ is electron charge, $\Omega$ is the cell volume, $E$ the energy, $w_{\vec{k}}$ is the Fermi-weight of each $k$-point, $e_\alpha$ are unit vectors along the three Cartesian directions, $\left|\psi_{n\vec{k}}\right\rangle$ is the cell-periodic part of the pseudo-wavefunction for band $n$ and k-point $k$, $q$ stands for the Bloch vector of an incident wave, $c$ and $v$ stand for conduction and valence bands, $\xi$ stands for eigenvalues of the corresponding bands respectively. The matrix elements on the right side of Eq. (1) capture the transitions allowed by symmetry and selection rules [32]. The real part of the dielectric tensor $\varepsilon_{\alpha\beta}^{(1)}$ is obtained by the usual Kramers-Kronig transformation [33]:

$$\varepsilon_{\alpha\beta}^{(1)}(E) = 1 + \frac{2}{\pi} P \int_0^\infty \frac{\varepsilon_{\alpha\beta}^{(2)}(E')E'}{(E')^2 - E^2 + i\eta} dE' \qquad (2)$$

where $P$ denotes the principle value, and $\eta$ is the complex shift parameter taken as 0.1. Moreover, as the dielectric function is a tensorial quantity, we use the crystallographic average of the dielectric function (written as $\varepsilon^{(1)}$ and $\varepsilon^{(2)}$), obtained by diagonalizing the dielectric tensor for each energy and averaging the diagonal elements.

Using, $\varepsilon^{(1)}$ and $\varepsilon^{(2)}$ the absorption coefficient $\alpha(E)$ is defined as:

$$\alpha(E) = \frac{2E}{\hbar c}\sqrt{\frac{\sqrt{\left(\varepsilon^{(1)}(E)\right)^2+\left(\varepsilon^{(2)}(E)\right)^2}-\left(\varepsilon^{(1)}(E)\right)}{2}} \qquad (3)$$

where $c$ is the speed of light.

Next, the SLME (η) is defined as the ratio of the maximum output power density ($P_{max}$) and the total incident solar energy density ($P_{in}$). $P_{max}$ is obtained by numerically maximizing the product of current density $J$ and voltage $V$.

$$\eta = \frac{P_{max}}{P_{in}} \qquad (4)$$

Assuming the solar cell at temperature $T$ behaves as an ideal diode and is illuminated under the photon flux $I_{sun}$, $J$ and $V$ follow the following equation:

$$J = J_{sc} - J_0\left(e^{\frac{eV}{kT}} - 1\right) \qquad (5)$$

where $e$ is the elementary charge, $V$ the potential over the absorber layer and $k$ is Boltzmann's constant. The first term is the short-circuit current density $J_{sc}$ given by:

$$J_{sc} = e\int_0^{\infty} a(E) I_{sun}(E) dE \qquad (6)$$

where and $a(E)$ is the photon absorptivity, $I_{sun}$ is the AM1.5G solar spectrum[34]. The $a(E)$ depends on the absorption coefficient ($\alpha$) (Eq. (3)) and thickness ($L$) of the material.

$$a(E) = 1 - e^{-2\alpha(E)L} \quad (7)$$

The coefficient of the second term in Eq. (5) is the reverse saturation current ($J_0$), which corresponds to the total (radiative and non-radiative) electron-hole recombination current at equilibrium in the dark :

$$J_0 = J_0^r + J_0^{nr} = \frac{J_0^r}{f_r} \quad (8)$$

Here, $f_r$ is defined as the fraction of the radiative recombination current. For the SLME, $f_r$ is approximated using:

$$f_r = e^{\left(\frac{E_g - E_g^{da}}{kT}\right)} \quad (9)$$

Where $E_g$ is the fundamental and $E_g^{da}$ is the direct allowed bandgap of a material.

Following the principle of detailed balance, the rates of emission and absorption through cell surfaces must be equal in equilibrium in the dark. Hence, $J_0^r$ can be calculated from the rate at which black-body photons from the surrounding thermal bath are absorbed through the front surface, given by:

$$J_0^r = e\pi \int_0^\infty a(E) I_{bb}(E,T) dE \quad (10)$$

where $I_{bb}$ is the black-body spectrum at temperature $T$. Both the solar spectrum $I_{sun}$ and black-body spectrum $I_{bb}$ are expressed in terms of the photon flux.

In order to maximize the power density, Eq. (1) can be re-written as:

$$\eta = \frac{P_{max}}{P_{in}} = \frac{\max\left\{\left(J_{sc} - J_0\left(e^{\frac{eV}{kT}} - 1\right)\right)V\right\}_V}{\int_0^\infty E I_{sun}(E) dE} \quad (11)$$

Therefore, the material-property related inputs in calculating the SLME are $\alpha(E), f_r, L$ and T. In this work, we assume material thickness ($L$) as 500 nm and temperature ($T$) as 300 K.

VASP uses a complex shift (CSHIFT) in the Kramers-Kronig relation to calculate the real part of the dielectric tensor, and also determines the corresponding imaginary part for consistency. This introduces a smoothening for both the real and imaginary part of the dielectric tensor, which translates to an earlier onset for the absorption spectrum. As this earlier onset can have a significant and unphysical influence on the calculated efficiency, we set the absorption to zero below the band gap. This is discussed in more detail in the supplementary information (Fig. S1, S2). Even with this correction, the SLME is still pushed slightly towards the SQ limit because of the increased onset produced by smoothening and cutting off the absorptivity. As this increase in efficiency does not lead to the elimination of potentially good photovoltaic materials, it is acceptable for our screening purposes. An example of the influence of removing the absorption below the band gap can be found in the supporting information (Fig. S1, S2). Note that the SLME uses an exponential function to model the fraction of radiative recombination as given in Eq. (6). As a consequence, the SLME quickly goes to zero as the difference between the direct allowed and the fundamental band becomes larger. This issue has been recently brought up by Bercx et al.[5] and Blank et al.[35] However, we consider the SLME as an appropriate metric for the initial screening of efficient materials for a thin film photovoltaic, which are generally direct bandgap in nature. Also note that for some materials, the SLME slightly exceeds the SQ limit of the corresponding band gap, which is related to the fact that by using a step function for the absorptivity *a(E)*, the SQ limit maximizes the reverse saturation current $J_0$. This is discussed in more detail in our previous work[5, 36]. For a selected set of materials, we performed $G_0W_0$ calculations[11, 37] with an ENCUTGW parameter (energy cutoff for response function) = 333.3 eV, 200 empty bands, and both with and without the inclusion of spin-orbit coupling (SOC).

We train a supervised machine learning (ML) classification model for predicting high-efficiency SLME (10% as a threshold) using classical force-field inspired descriptors (CFID)[23] and mainly gradient boosting decision trees (GBDT) algorithm[38]. The CFID gives a unique representation of a material using structural (such as radial, angle and dihedral distributions), chemical (such as average electronegativity, average heat of fusion for constituting elements), and charge descriptors (average of radial charge around the nucleus of each atom). For an arbitrary material, CFID generates 1557 descriptors. The principal idea behind the GBDT algorithm is to build the new base learners to be maximally correlated with the negative gradient of the loss function, associated with the whole ensemble. We categorize the SLME data of all materials as 0 or 1 depending on whether they have SLME ≥ 10%. Hence, the ML model is simply a binary classification model. As the number of materials with SLME <10 % outnumber the number of materials with SLME ≥ 10% in our dataset, the baseline of the ML model is predicting the material has a low-SLME. As a standard practice, we divide the whole SLME dataset using 90%-10% split [39, 40]. We train ML on 90% data and test the ML on 10% data to evaluate performance. Specifically, we evaluate the performance of ML models based on the area under curve (AUC) for receiver operating characteristics (ROC)[41, 42].

For a classifier and an instance, there could be four possible outcomes. If the instance is positive and classified (with the classifier model) as positive, it is counted as True Positive (TP), if it is classified negative, it is counted as False Negative (FN). If the instance is negative and it is classified as negative, it is counted as a True Negative (TN); if it is classified as positive, it is counted as a False Positive (FP). So, for example, if the SLME of a material is 10% or more and the classifier ML model actually predicts it to have high-SLME, then the material is a TP and so on. Now, the True Positive Rate (TPR) is the ratio of positives correctly classified (TP) and total positives (TP+FN), while false positive rate (FPR) is the negatives incorrectly classified (TN) with total negatives (TN+FP). Plotting TPR against FPR in a ROC curve depicts relative tradeoffs between benefits (true positives) and costs (false positives) suggesting how good the model can distinguish between two classes (e.g if a material has a high or low SLME). A good model can accurately distinguish between the two, whereas, a poor model will have difficulties in distinguishing between the two. In our work, a classifier model predicts probabilities (between value 0.0 and 1.0) for each material that the material is a high or low SLME. Now a particular threshold probability would correspond to a TPR and an FPR and if we plot all such points for all possible thresholds, we obtain a ROC curve. An area under curve (AUC) of 1.0 signifies a perfect model, while AUC with 0.5 denotes random-guessing. More details about ROC curve can be found in Ref.[41, 42].

We first train classification models with default parameters using decision-trees, random-forest, k-nearest neighbor, multi-layer perceptron, and gradient boosting models implemented in scikit-learn[43] package, and also GBDT implemented in XGBoost[44] and LightGBM[38] packages. The GBDT implementations have slightly different implementations in scikit-learn, XGBoost and LightGBM. As the LightGBM's GBDT gives overall high accuracy (discussed later), we further tune the hyperparameters (mainly learning rate, number of trees and maximum number of leaves) in the LightGBM using a five-fold cross-validation grid-search on the 90% training set. The model hyperparameters are provided in the supplementary information. Using the best model found during the grid-search, we test the model on 10% held set and report the ROC and AUC. We use this model on a large set of materials to quickly pre-screen materials with high-SLME. Details of the SLME-code used for running the high-throughput workflow, calculating the SLME, and training the machine learning model is available at: https://github.com/usnistgov/jarvis.

## Results and discussion

The SLME can be considered as the theoretical maximum of the photo-conversion efficiency of a single p-n junction solar cell[45]. We calculate the SLME ($\eta$) for an absorber layer with thickness 500 nm and at 300 K for all the materials in our database for which the frequency-dependent dielectric function is available. Out of 30000 materials in JARVIS-DFT, 12881 materials have TBmBJ bandgap and frequency-dependent dielectric function data and the database is still growing. Considering only non-metallic systems leads to 5097 materials for the calculation of the SLME values. Using 10 % as a threshold, 1997 solar cell material candidates remain, which significantly expands the list of known solar materials. The list of candidates includes several already known materials[4] such as CdTe, GaAs, $CuInSe_2$, $CuGaSe_2$, $ZnSnP_2$, $CdSnP_2$ and $CH_3NH_3PbI_3$ as well as many new ones (discussed later). To benchmark the screening workflow, we compare our SLME for five direct-bandgap compounds with respect to experimental data in Table. S1. Note that experimental samples generally have plenty of defects which are completely absent in our theoretical calculations. We find that for all of the materials, the calculated SLME is higher than the experimental efficiency. As the SLME provides an upper limit for the efficiency and can hence be used to eliminate materials which are found to have a low theoretical maximum efficiency, it is gratifying to see that none of the materials found to be suitable in experiment would be removed based on the SLME metric. The calculated mean absolute deviation (MAD) between TBmBJ and experiments is 4.80-8.30 %. Furthermore, we also compare 10 compounds' SLME obtained with the TBmBJ and $G_0W_0$ method[4] in Table S2. There is an overall decreasing trend in SLME for $G_0W_0$ data compared to TBmBJ. The calculated mean absolute deviation (MAD) between TBmBJ and $G_0W_0$ is 5.21 %, which is reasonable considering the number of materials investigated. The MAD between theoretical $G_0W_0$ is smaller than experimental data signifying better comparison among theoretical results.

Figure 1 shows the SLME and property distribution of the investigated materials. In Fig. 1a we observe that our criterion on the SLME eliminates more than 50 % of the materials. For the candidate materials, we analyze their characteristics to further identify interesting trends. In Fig. 1b we plot the SLME versus the bandgap and observe that, although a material could be deemed suitable as a solar cell material using traditional approaches like SQ, the SLME shows efficiency values over a wide range for materials with similar bandgaps. This indicates that the SLME is a stricter selection metric than the SQ limit, which is based solely on the band gap. This was previously demonstrated by Yu et al.[4], but for a smaller set of materials. To further elucidate the SLME and bandgap relationship, Fig. 1c shows a colormap of the SLME values versus the direct and indirect bandgap. Clearly, direct-bandgap materials close to 1.1 eV have a high SLME, which can be explained by the SLME's origin from the SQ-formalism. Note that the SLME tends to underestimate the efficiency of materials with a large difference between the fundamental an optical band gap. Next, the calculated electron effective mass ($m^*$) is included in the screening process, by eliminating materials which have an electron effective mass values close or higher than 1.0 $m_0$ where $m_0$ is the mass of a free electron. The effective masses at 300 K were determined using an approach based on the Boltzmann-transport equation as implemented in the BoltzTrap[46] code, and are plotted in Fig. 1d. The effective mass plays an important role in designing solar-cells even if the material is highly absorbing because heavy charge-carriers are an indication of low efficiency. Note that as we are not aware of a metric combining absorption coefficient and effective mass, we simply perform a secondary screening solely based on the effective mass values.

Due to the recent explosion in low-dimensional materials research[47], it is interesting to see how many of the candidate materials belong to this class. Fig. 1e demonstrates that while most of the predicted materials are 3D, there are significant contributions from low dimensional materials as well. The number of low dimensional materials are determined by using the combined lattice-constant and data-mining approaches[21]. Note that the dimensionality is considered to be reduced if there exists vdW bonding in one/two/three directions. Finding low dimensional materials can be of great interest because they allow for high carrier mobility and easy thin-film fabrication. We mainly focus on layered materials i.e. 2D materials in their bulk forms. As there are several initiatives to build solar panels around curved shapes/architectures using flexible low-dimensional materials, the low dimensional materials predicted here could be of significant technological interest. In Fig. 1f, we see that most of the high-efficiency materials are ternary, which is consistent with known thin-film materials[4] such as chalcopyrite $CuInSe_2$ and $AgCuSe_2$. However, multicomponent systems can be difficult to fabricate experimentally, and in those cases, the list of less complicated compositions could be of interest to experimentalists. From Fig. 1g to Fig. 1i, it is clear that the efficiency of a material is only weakly correlated with the crystal system, compositional prototypes or space-groups, indicating that a simple structural screening is not sufficient, and detailed electronic structure calculations are essential for accurately predicting the efficiency of absorber materials. Although solar cell materials can belong to a wide variety of crystalline systems, there are some large fractions of suitable materials for space groups which correspond to those of well-known solar cell materials such as chalcopyrites (space group 122) and perovskites (space group 221). Apart from 122 and 221, other space groups with a large fraction of high potential materials are 225, 166, 12, 216, 62 and 194. From a technological synthesis perspective, a particular crystal system could be favorable to experimentalists such as the case of perovskite solar cells[48].

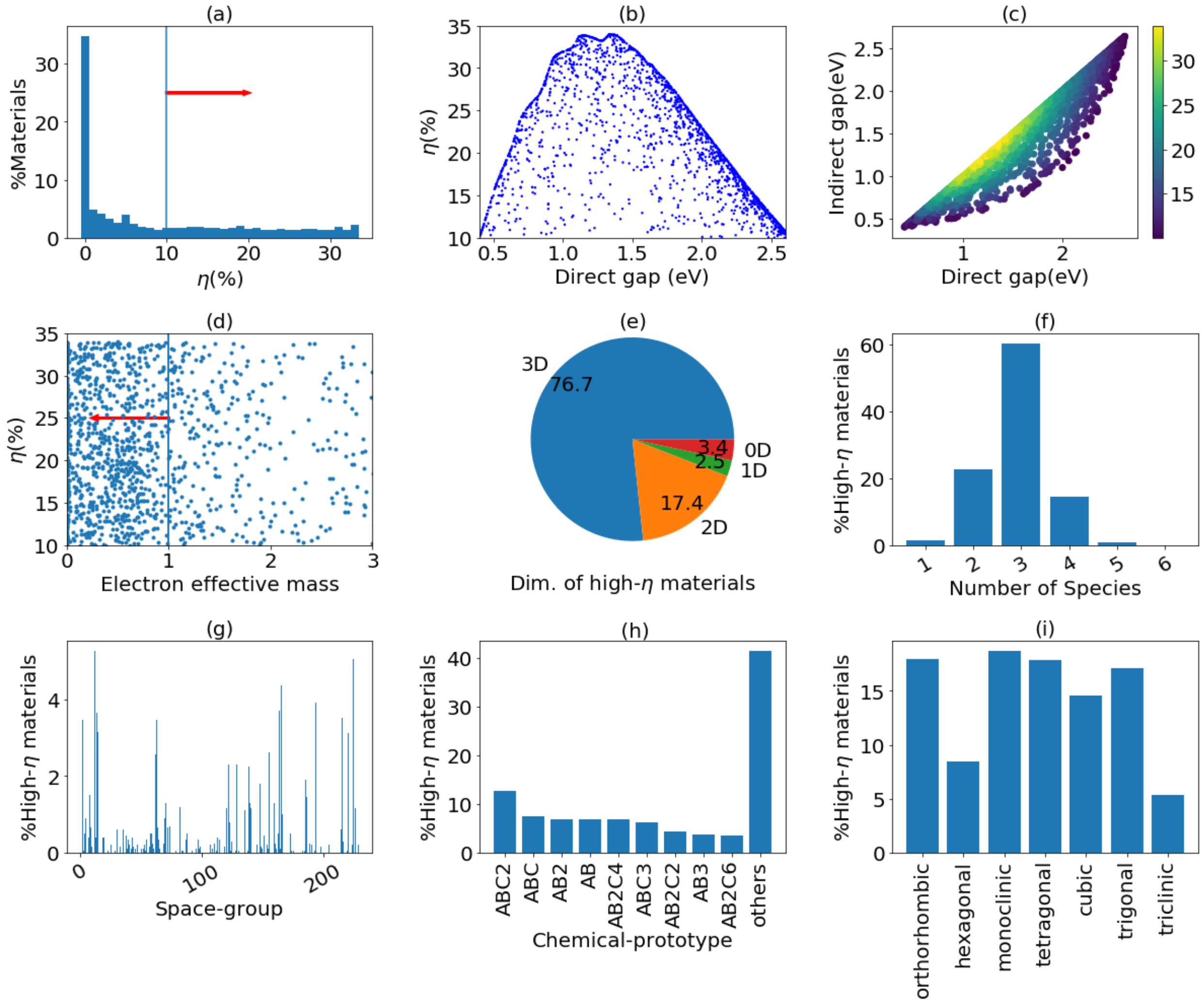

*Fig. 1 Summary of SLME data. a) SLME-distribution of all the materials in the database, b) SLME (η) vs TBmBJ fundamental bandgap for high-SLME materials, c) colormap of SLME values with*

*the direct versus indirect bandgaps for high-SLME materials, d) SLME vs average effective mass of electrons, e) dimensionality distribution in terms of 3D-bulk, 2D-bulk, 1D-bulk and 0D-bulk materials in the database, f) number of species distribution for high SLME materials, g) space-group distributions (1-230) for high-SLME materials, h and i) compositional-prototype and crystal-system distributions for high-SLME materials.*

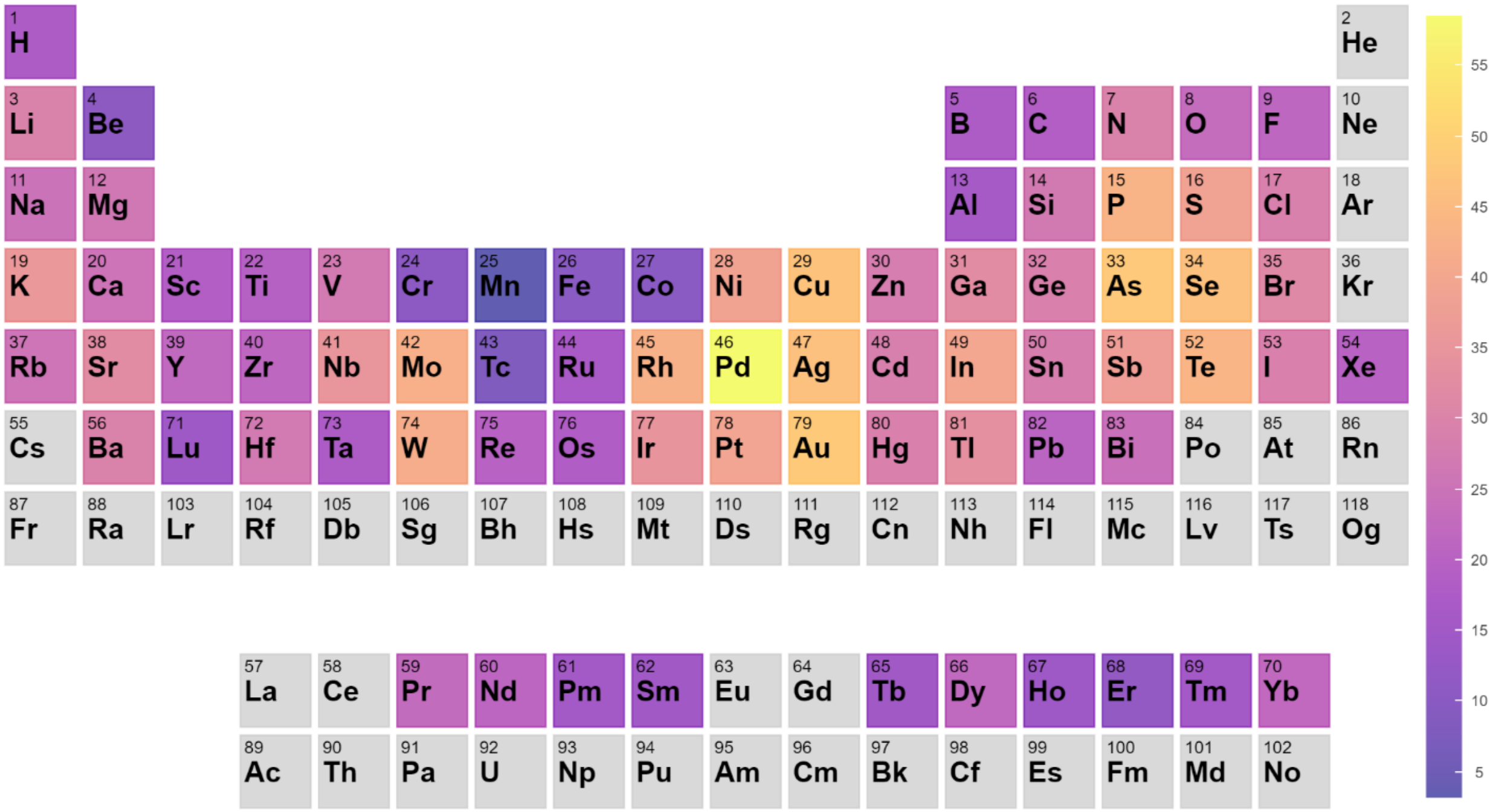

*Fig. 2 Periodic table trends of high SLME materials. The elements in a material are weighed 1 or 0 if the material has high or low-SLME. Then the probability of finding the element in a high-SLME material is calculated.*

While it is interesting to compare the SLME with the crystallographic information in the previous figure, it would also be beneficial to see which elements from the periodic table contribute most to the high-efficiency materials. Generally, there is no established way of identifying high-efficiency solar cell materials based on just elements, but such periodic table trends can be used as an initial guideline for material design. In order to understand the elemental contributions, we weigh an element in a material one or zero depending on whether the material has an SLME above 10 % or not. After such weighing for all the materials in our database, we calculate the probability that an element is part of a high-efficient SLME material. Suppose there are x number of Se-containing materials and y of them have SLME greater equal to 10%, then the percentage probability (p) for Se is calculated using the formula: $p = \frac{y}{x} \times 100\%$ The results in Fig. 2 indicate that transition metals and chalcogenides such as Cu-Ag-Au, Mo-W, Rh-Ni-Pt, Ga-In, Tl, P, As, B, and K are the main constituent elements of high SLME materials. This is again in agreement with widely known efficient chalcopyrite materials[4] such as Copper Indium Gallium Selenide (CIGS). Remarkably, the combination of transition metals and chalcogenides led to commonly known transition metal chalcogenide (TMD) materials which are of great interest for 2D material applications. Note that although the TBmBJ formalism can be safely used to calculate the properties of low-dimensional materials in their bulk form, the inclusion of excitonic effects is critically important for calculating accurate absorption coefficients of monolayer materials. Hence, the focus of this work is on bulk periodic materials. However, having the predicted dimensionality is important to know whether it is possible to exfoliate the material in one/two/three directions.

Next, we present a few screening examples for solar cells. First, as mentioned above, most of the chalcopyrites have space group 122, so we screen materials with space-group 122 and SLME $\geq$ 10 %, which results in 44 materials. Further screening based on reduced effective masses < 1.0 $m_0$ and energy above the convex hull < 0.1 eV/atom leads to materials such as $MgGeP_2$ (JVASP-8813), $ZnSiAs_2$ (JVASP-2256), and $AlCuS_2$ (JVASP-2397). We are not aware of any previous literature which has reported these materials as potential photovoltaic materials. Similar searches for finding perovskites with SLME $\geq$ 10 % and space-group 221 results in materials such as $TaTlO_3$ (JVASP-41734) and $TiSnO_3$ (JVASP-35817). Some other classes of high-SLME materials are chalcogenides such as: $XY_2Z_4$(X=Zn, Ba,Sr, ; Y=In, Ga; Z=Te, Se), XY (X=Ga, Zn, Sb, Cd; Y=O,Te,Se), $XYTe_2$( X=Rb, Na, Ag, Y=Y, Al, Ga), $XPS_3$(X=K, Sn, Rb, Tl), $WX_2$(X=Se, Te, N), $X_2Te_5$(X=In, Ga, Al), $XCu_3Te_4$ (X=V, Ta), halides such as: $GeKX_3$(X=Cl, Br), $K_2X_4Y$(X=Br, Cl, F; Y=Pd, Pt), $PdX_2Y_6$ (X=Rb,Se, Y=O,Cl) and many other distinct classes such as $Sb_2XO_8$(X=Mg, Zn), $Sb_2Mg_2X$(X=Ca, Sr, Ba), $Sb_2K_2X$(X=Rb, Cu, Ag, Au), $XO_2$(X=Ti, Ni), SiXY(X=Pd, Pt; Y=Ti, Zr). In addition, there are also numerous compounds with unique chemical prototypes. Hence, our screening approach predicts several orders of magnitude new compounds and new classes that can be of immense importance to the solar-cell community.

Finally, we perform a screening of all the 2D-bulk materials with a high SLME, low reduced effective mass (< 1.0 $m_0$), and energy above the convex hull < 0.1 eV/atom. In our previous work[14], we found that TBmBJ accurately predicted the dielectric function of 2D-bulk materials such as $MoS_2$ and $SnSe_2$. The dimensionalities of the bulk materials were determined with lattice-parameter and data-mining approaches[21] as mentioned above. We identified at least 58 potential 2D-bulk materials based on our screening criteria (Table 1). In order to further analyze the TBmBJ accuracy for these materials, we performed $G_0W_0$ calculations for five of the 58 materials. The comparison of the TBmBJ and $G_0W_0$ results is shown in Table. S2. We only investigated five materials because of the enormous computational time necessary for running $G_0W_0$ and $G_0W_0$+SOC calculations. The computational times are given in the supplementary information (Table. S3). We find that the MAD between the TBmBJ and $G_0W_0$ for band gaps and SLME are 0.22 and 3.23 % respectively. The MAD for SLME further drops by the inclusion of spin-orbit coupling (SOC) in the $G_0W_0$ calculations. These low computationally-derived MAD values confirm the high-performance of the candidate materials. Note that SOC is not considered for the TBmBJ calculations. In the future, we would like to carry out calculations for more materials among the 1997 candidates to further carry out the benchmarking analysis.

*Table 1: The JARVIS-ID (JVASP), chemical formula, crystallographic space-group, SLME, TBmBJ fundamental gap ($E_g$), average electron effective mass ($m^*/m_0$) of all the 2D-bulk layered materials with high SLME (>10 %), low effective mass (<1.0$m_0$) and energy-above the convex hull (0.1 eV/atom) are shown as an example of screening.*

| JID | Formula | Spg. | $E_g$ | SLME | $m^*/m_0$ | JID | Formula | Spg. | $E_g$ | SLME | $m^*/m_0$ |
|---|---|---|---|---|---|---|---|---|---|---|---|
| **8781** | BiTeBr | 156 | 1.9 | 25.2 | 0.3 | **13064** | $Tl_2Au_4S_3$ | 59 | 1.6 | 31.2 | 0.64 |
| **26802** | $AgBiSCl_2$ | 63 | 1.6 | 31.3 | 0.9 | **14351** | $Rb_2TeI_6$ | 128 | 1.8 | 18.5 | 0.5 |
| **179** | $GeI_2$ | 164 | 2.5 | 12 | 0.72 | **131** | $SnS_2$ | 164 | 2.1 | 10.6 | 0.57 |
| **54** | $MoS_2$ | 194 | 1.3 | 18.6 | 0.59 | **51** | $MoS_2$ | 160 | 1.3 | 21.1 | 0.79 |
| **5644** | GeAsSe | 52 | 2.1 | 19.6 | 0.45 | **4358** | GaSe | 187 | 2.1 | 21.4 | 0.12 |
| **60** | $MoTe_2$ | 194 | 1 | 29 | 0.49 | **57** | $MoSe_2$ | 194 | 1.3 | 27.6 | 0.53 |
| **122** | $SnSe_2$ | 164 | 1.1 | 17 | 0.11 | **4216** | SiAs | 12 | 1.6 | 26.1 | 0.41 |
| **299** | SnSe | 62 | 1.3 | 24 | 0.24 | **81** | GaSe | 194 | 2.1 | 20 | 0.12 |
| **231** | $MoSe_2$ | 160 | 1.3 | 30.4 | 0.74 | **5053** | $InAg(PSe_3)_2$ | 163 | 1.4 | 32 | 0.16 |
| **4630** | $TlPt_2S_3$ | 164 | 1.3 | 32.7 | 0.61 | **5146** | $InAg(PS_3)_2$ | 163 | 1.9 | 19.5 | 0.26 |
| **29420** | $AgBiSCl_2$ | 63 | 1.6 | 31.3 | 0.92 | **5176** | CuBr | 129 | 1.9 | 25.6 | 0.29 |
| **29475** | SnS | 63 | 1.2 | 18.8 | 0.63 | **5215** | $Bi_2Se_3$ | 62 | 1.4 | 32.3 | 0.32 |
| **29566** | $HgI_2$ | 137 | 2.1 | 20.8 | 0.22 | **5224** | $HgI_2$ | 137 | 2 | 22 | 0.2 |
| **29640** | $SnPSe_3$ | 14 | 1.7 | 24.3 | 0.88 | **5269** | BiSI | 62 | 2.4 | 12.3 | 0.98 |
| **29801** | $Te(HO_2)_2$ | 14 | 2.4 | 14.5 | 0.66 | **4636** | $Mg(AlSe_2)_2$ | 166 | 2.2 | 18.1 | 0.22 |
| **29802** | $Te(HO_2)_2$ | 7 | 2.4 | 14.3 | 0.69 | **290** | $SnS_2$ | 186 | 2 | 16.7 | 0.37 |
| **29874** | $AgSbS_2$ | 15 | 1.2 | 29.2 | 0.32 | **3849** | AuI | 138 | 2.1 | 11.4 | 0.54 |
| **29884** | $CdInGaS_4$ | 164 | 1.5 | 31.5 | 0.14 | **3414** | InS | 58 | 1.8 | 10.7 | 0.58 |
| **13856** | GaSe | 160 | 2.1 | 20.7 | 0.14 | **1639** | $CdHgO_2$ | 12 | 1.6 | 30.2 | 0.26 |
| **14038** | $Hg_2IO$ | 15 | 1.7 | 23.4 | 0.48 | **10107** | $Tl_2GeS_3$ | 2 | 2.2 | 12.4 | 0.61 |
| **30064** | $Ag_2H_2IOF$ | 4 | 2.6 | 11.4 | 0.34 | **51** | $MoS_2$ | 160 | 1.3 | 21.1 | 0.79 |
| **30452** | $B_2S_3$ | 167 | 2.3 | 13.7 | 0.94 | **60** | $MoTe_2$ | 194 | 1 | 29 | 0.49 |
| **30460** | $BiS_2$ | 12 | 1.9 | 19.1 | 0.55 | **29284** | $Sn(PS_3)_2$ | 146 | 1.8 | 23.6 | 0.54 |
| **30494** | SnBrCl | 129 | 2 | 23.9 | 0.45 | **29294** | InTeCl | 14 | 2.5 | 10.5 | 0.23 |
| **22637** | $Cd(InSe_2)_2$ | 111 | 2.1 | 20.8 | 0.14 | **29359** | GaTe | 12 | 1.7 | 27.8 | 0.72 |
| **7785** | SnS | 63 | 1.6 | 14.7 | 0.62 | **8490** | InSe | 12 | 1.5 | 18.8 | 0.72 |
| **13003** | InTeBr | 14 | 2.4 | 14.1 | 0.41 | **8670** | SbTeI | 2 | 1.3 | 22.1 | 0.74 |
| **28369** | PbS | 63 | 1.9 | 25.7 | 0.81 | **4026** | BiTeCl | 186 | 1.8 | 26.5 | 0.25 |
| **9754** | $Tl_2Sn(AsS_3)_2$ | 147 | 1.7 | 26.4 | 0.42 | **1963** | BiTeI | 156 | 1.7 | 29.5 | 0.34 |

*Table 2: Bandgap and SLME properties of a selection of materials from Table 1 with TBmBJ and $G_0W_0$ methods in DFT to evaluate uncertainty in predictions. Here $E_g$ denotes the bandgap in eV and $\eta$ the calculated SLME in percentage.*

| **Materials** | **JID** | **$E_g$ (TBmBJ)** | **$E_g$ ($G_0W_0$)** | **$E_g$ ($G_0W_0$+SOC)** | **$\eta$ (TBmBJ)** | **$\eta$ ($G_0W_0$)** | **$\eta$ ($G_0W_0$+SOC)** |
|---|---|---|---|---|---|---|---|
| **CuBr** | 5176 | 1.9 | 2.01 | 2.09 | 25.6 | 22.74 | 21.04 |
| **AuI** | 3849 | 2.1 | 2.34 | 2.20 | 11.4 | 8.83 | 11.86 |
| **SiAs** | 4216 | 1.6 | 1.36 | 1.33 | 26.1 | 23.85 | 23.20 |
| **BiTeBr** | 8781 | 1.90 | 1.52 | 0.79 | 25.2 | 32.15 | 26.11 |
| **$TlPt_2S_3$** | 4630 | 1.30 | 1.45 | 1.35 | 32.70 | 30.99 | - |
| **MAD** | - | - | 0.22 | 0.34 | - | 3.23 | 2.21 |

Recently Walsh et al.[19] argued that the size of the design space of possible materials can be on the order of $10^{100}$. Carrying out high-level DFT calculations for materials on this scale is an impossible task due to the associated computational cost. Hence, we train a machine learning model[49] which can help in the screening process. Based on the SLME data, we classify materials in two classes: high (SLME≥10 %) and low (SLME<10 %) efficiency materials. In order to convert all the crystallographic information to computational fingerprints, we use the classical force-field inspired descriptors (CFID). We first train classification models with default parameters using decision-trees, random-forest, k-nearest neighbor, multi-layer perceptron, and gradient boosting models implemented in scikit-learn package, and also GBDT implemented in XGBoost and LightGBM packages. As a standard practice, we use train-test split (90 %:10 %) [39, 40] , five-fold cross-validation [50] and examining area under curve (AUC) for receiver operation characteristics (ROC)-curves on the 10 % held set (as shown in Table. 3).

*Table 3. Initial comparison of ML classification techniques using decision-trees (DT), random-forest (RF), k-nearest neighbor (KNN), multi-layer perceptron (MLP), GBDT implemented in scikit-learn package (SK-GB), GBDT in XGBoost (XGB) and GBDT in LightGBM (LGB).*

| **Model** | **DT** | **RF** | **KNN** | **MLP** | **SK-GB** | **XGB** | **LGB** |
|---|---|---|---|---|---|---|---|
| **AUC** | 0.67 | 0.79 | 0.77 | 0.80 | 0.84 | 0.84 | 0.87 |

Evidently, the LGB model already performs very well with default parameters only. We further tune LGB hyperparameters such as number of estimators, number of leaves and learning rate using a five-fold cross-validation grid-search. Using the best model of the grid-search we predict the ROC of the 10% held set (shown in Fig. 3) to give an AUC of 0.90. As AUC 1.0 suggests a perfect model, a ROC area of 0.90 suggests a highly accurate model. The model is publicly available at JARVIS-ML (https://www.ctcms.nist.gov/jarvisml/ ) to quickly predict whether or not the material will have a high SLME. To obtain possible materials, we use the large crystallographic databases such as AFLOW[24], Materials-project (MP)[25], OQMD[26] and COD[28]. We convert the crystal structure into CFID descriptors for 639262 materials from AFLOW, 82125 materials from MP, 360802 materials from OQMD, and 111783 materials from the COD database. Out of 1193972, we find 669051 materials with unique chemical compositions and spacegroups and 306469 with unique chemical compositions only. After applying the trained classification model on these materials, we pre-screen 8970 materials, which can be used to narrow down and prioritize future DFT calculations. Out of these 8970 materials, 6342 have unique chemical compositions, while the rest can have the same chemical compositions but different spacegroups. The list of materials pre-screened using ML models is also provided in the supporting information. The properties of these materials will be determined using OptB88vdW and TBmBJ calculations within the JARVIS-DFT workflow. Hence, based on the ML model, we can optimize the DFT screening process. As this feedback loop keeps learning new data, we expect the ML model to continuously improve its accuracy in a controlled and systematic way.

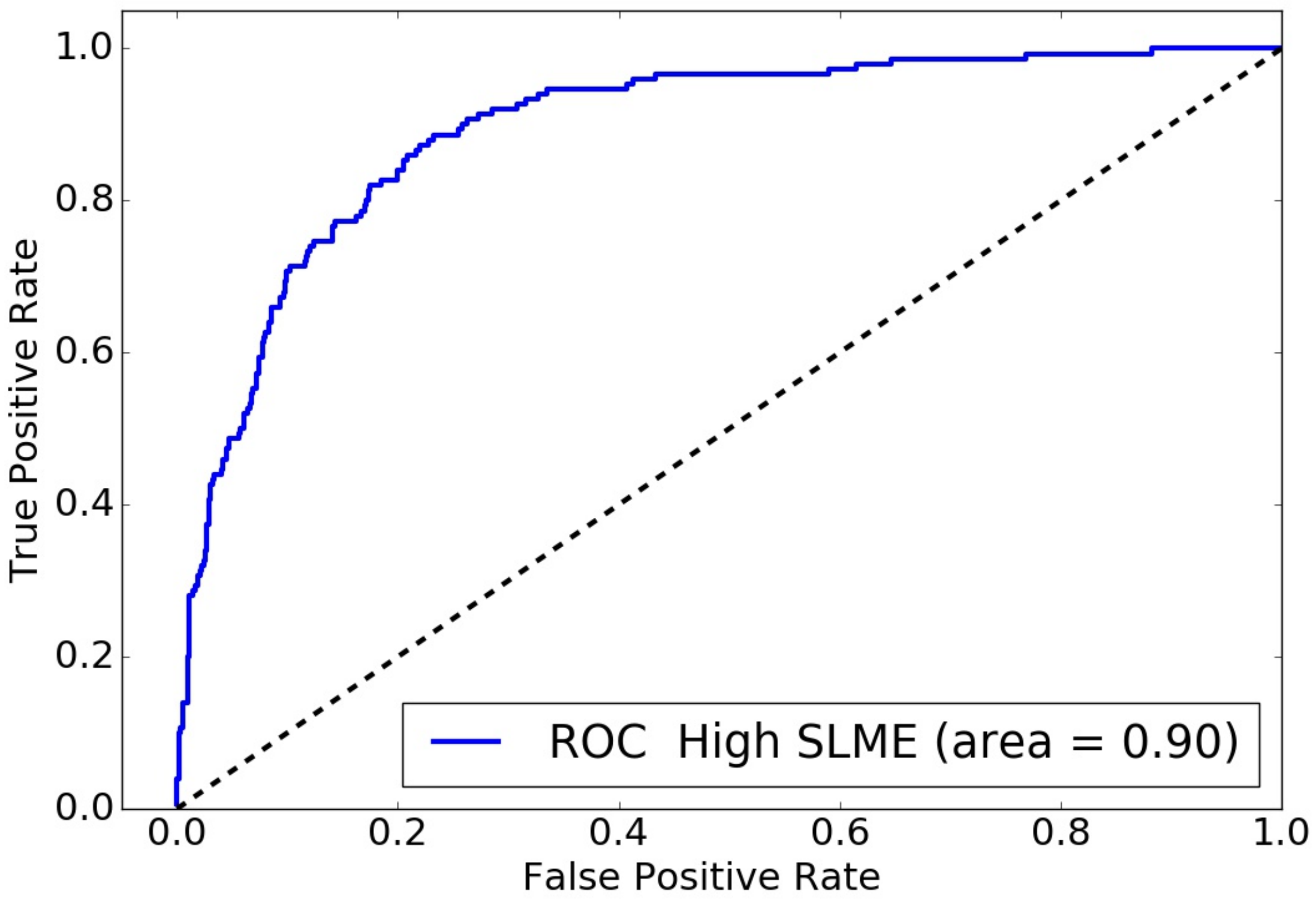

*Fig. 3 Classification receiver-operating characteristic (ROC) curve for high-SLME materials. The dotted line shows the random guessing line with an area under curve 0.5.*

## Conclusions

In summary, we have presented the results of a combined density functional theory high-throughput screening and machine learning approach for identifying promising solar cell materials based on the spectroscopy limited maximum efficiency. Using frequency-dependent dielectric function data obtained with the meta-GGA TBmBJ formalism drastically increases the volume of materials data which can be investigated with high accuracy. Additionally, we use the effective carrier mass and energy above the convex hull to further screen candidate materials. Our analysis reveals several trends for high-efficiency solar materials starting from crystallographic information to chemical constituents and identifies 58 potential 2D-bulk solar cell materials with high potential as thin-film solar-cell materials. Finally, we have trained a machine learning classification model with the SLME data which can quickly predict whether a material will have an SLME above 10%. We believe the data, tools and the methodology for identifying solar cell materials provide a complete suite to accelerate the discovery of photovoltaic materials and can have a significant impact on the next-generation of materials design.

## Supporting information

The Supporting Information is available free of charge on the ACS Publications website at…

- The figshare link contains data generated by the density functional theory and machine learning (ML) study,
- The ML hyperparameters for the trained model,
- Comparison of computational cost for TBmBJ, G0W0 and G0W0+SOC methods,
- Effect of complex shift (CSHIFT) and setting absorption coefficient zero below bandgap on SLME is shown.

# Supplementary info: Accelerated Discovery of Efficient Solar-cell Materials using Quantum and Machine-learning Methods

Kamal Choudhary[1], Marnik Bercx[2], Jie Jiang[3], Ruth Pachter[3], Dirk Lamoen[2], and Francesca Tavazza[1]

1 Materials Science and Engineering Division, National Institute of Standards and Technology, Gaithersburg, Maryland 20899, USA.

2 EMAT, Department of Physics, University of Antwerp, Groenenborgerlaan 171, 2020 Antwerp, Belgium.

3 Materials Directorate, Air Force Research Laboratory, Wright–Patterson Air Force Base, Ohio 45433, USA.

**Figshare datalink**: We provide the data generated in this study at **https://doi.org/10.6084/m9.figshare.8218940 .**

*Table. S1 Comparison of experimental efficiency with DFT-TBmBJ SLME data. Details of individual materials can be found at corresponding JARVIS-ID webpages, for example,* https://www.ctcms.nist.gov/~knc6/jsmol/JVASP-1174.html *for JARVIS-ID-JVASP-1174. The experimental data were obtained from Kasap*[48] *and Green et al.*[49].

| Materials | Expt. Eff. (%) | η (TBmBJ) (%) | JARVIS-ID |
|---|---|---|---|
| GaAs | 24.0-29.7 | 33.64 | JVASP-1174 |
| CdTe | 15.0-21.4 | 28.17 | JVASP-7757 |
| InP | 21.0-24.7 | 33.84 | JVASP-266 |
| $CuInSe_2$ | 12.0-13.0 | 21.35 | JVASP-8554 |
| Cubic-$CH_3NH_3PbI_3$ | 16.0-20.3 | 20.82 | JVASP-7112 |
| **MAD** | - | 4.80-8.30 | |

*Table S2: Comparison of SLME using TBmBJ and GW methods for a few materials used in Yu et al's work*[4].

| Materials | ICSD-ID | JARVIS-ID | η ($G_0W_0$) | η (TBmBJ) |
|---|---|---|---|---|
| $CuInS_2$ | 656271 | JVASP-8546 | 23.1 | 23.7 |
| $CuInTe_2$ | 658015 | JVASP-3495 | 28 | 32.2 |
| $CuGaSe_2$ | 627528 | JVASP-8071 | 26.6 | 33.8 |
| $AgInTe_2$ | 605485 | JVASP-8532 | 26.4 | 31.6 |
| $AgIn_5Te_8$ | 151871 | JVASP-12861 | 26.3 | 32.83 |
| $CuGaTe_2$ | 656165 | JVASP-2295 | 24.8 | 32.17 |
| $CuInSe_2$ | 602951 | JVASP-8554 | 22.1 | 21.35 |
| $CuBSe_2$ | 613591 | JVASP-12878 | 20.6 | 20.45 |
| $AgInS_2$ | 656317 | JVASP-3420 | 19.7 | 30.45 |
| $AgIn_5Se_8$ | 35597 | JVASP-12862 | 22.2 | 31.5 |
| **MAD** | - | - | - | 5.2 |

*Table S3: Comparison of computational cost (Total CPU time in seconds * number of cores/3600) for each calculation.*

| Materials | TBmBJ | G0W0 | G0W0+SOC |
|---|---|---|---|
| **CuBr** | 31.64 | 206.25 | 4381.2 |
| **SiAs** | 317.3911 | 7427.833 | 24615 |
| **TlPtS** | 510.0889 | 1607.4 | 12758.34 |
| **BiTeBr** | 2.782222 | 131.4444 | 5562.667 |
| **AuI** | 124.8933 | 1487.5 | 9645.044 |

**The ML model parameters**: n_estimators (5000), learning_rate (0.1), max_depth (50), boosting_type('gbdt'), min_child_samples (20), min_child_weight (0.001), min_split_gain (0.0), num_leaves (100), reg_alpha (0.0), reg_lambda (0.0), subsample (1.0), subsample_for_bin (200000, subsample_freq=1). All other parameters are taken as default hyperparameters. More details about these hyperparameters can be found in the LightGBM package documentation (https://lightgbm.readthedocs.io/en/latest/Parameters.html).

**Effect of CHSIFT and setting absorption zero below bandgap:**

Using the standard value for the complex shift in the Kramers-Kronig transformation (CSHIFT) in VASP leads to a broadening of the dielectric function for both the imaginary and real part. The most important influence on the absorption spectrum is due to the broadening of the imaginary part, as this changes the onset of the absorption coefficient. In Figure S1 we show the imaginary part of the dielectric function $\varepsilon^{(2)}$, as well as the absorption coefficient. We can see that because of the broadening introduced by the CSHIFT, the imaginary part of the dielectric function is not zero below the band gap. Even though $\varepsilon^{(2)}$ is small, it has a significant influence on the absorption coefficient, which has a value up to approximately $10^5$ $\mathrm{m}^{-1}$ below the band gap.

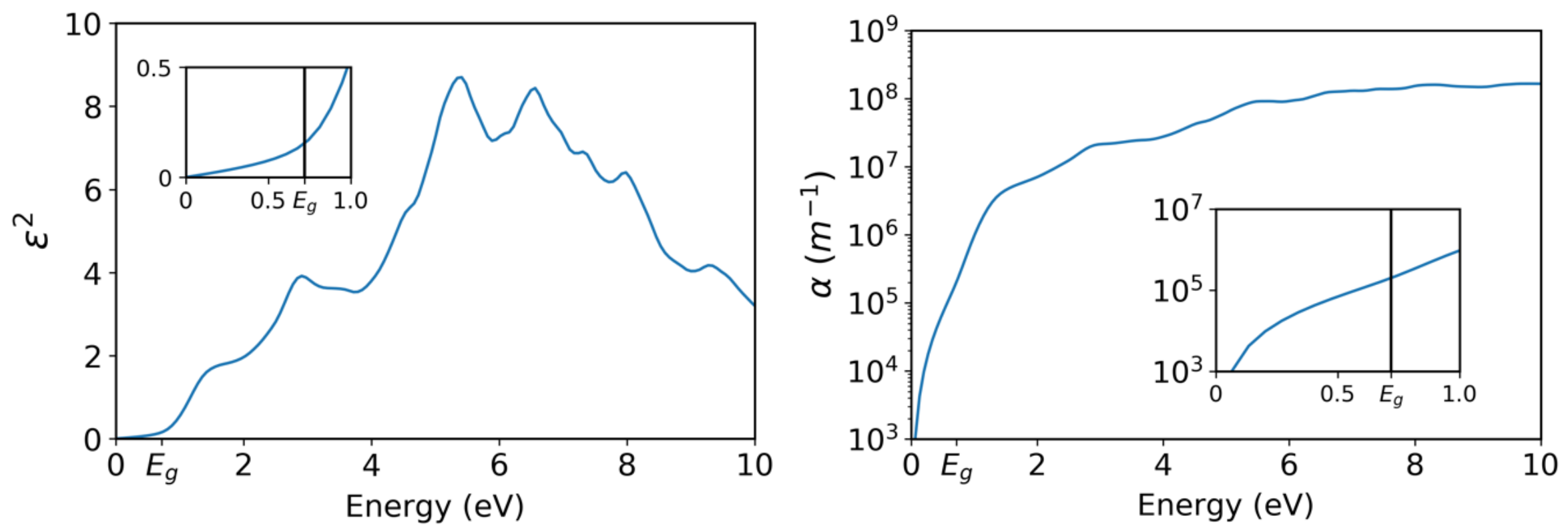

*Figure S1: Imaginary part of the dielectric function $\varepsilon^{(2)}$ (left) and absorption coefficient $\alpha$ for CuAu-like $CuInSe_2$. In both figures, the inset focuses on the energy range 0-1 eV in order to better visualize the onset below the calculated band gap of $CuInSe_2$ (0.72 eV).*

As can be seen in Fig. S2, this additional absorption severely affects the SLME for all thicknesses. Without setting the absorption coefficient to zero for energies below the band gap, the calculated SLME does not exceed 3% for any thickness. When we do set $\alpha(E) = 0$ for $E < E_g$, we obtain more reasonable values for the maximized efficiency, and the SLME converges to the Shockley-Queisser value of the corresponding band gap (0.72 eV).

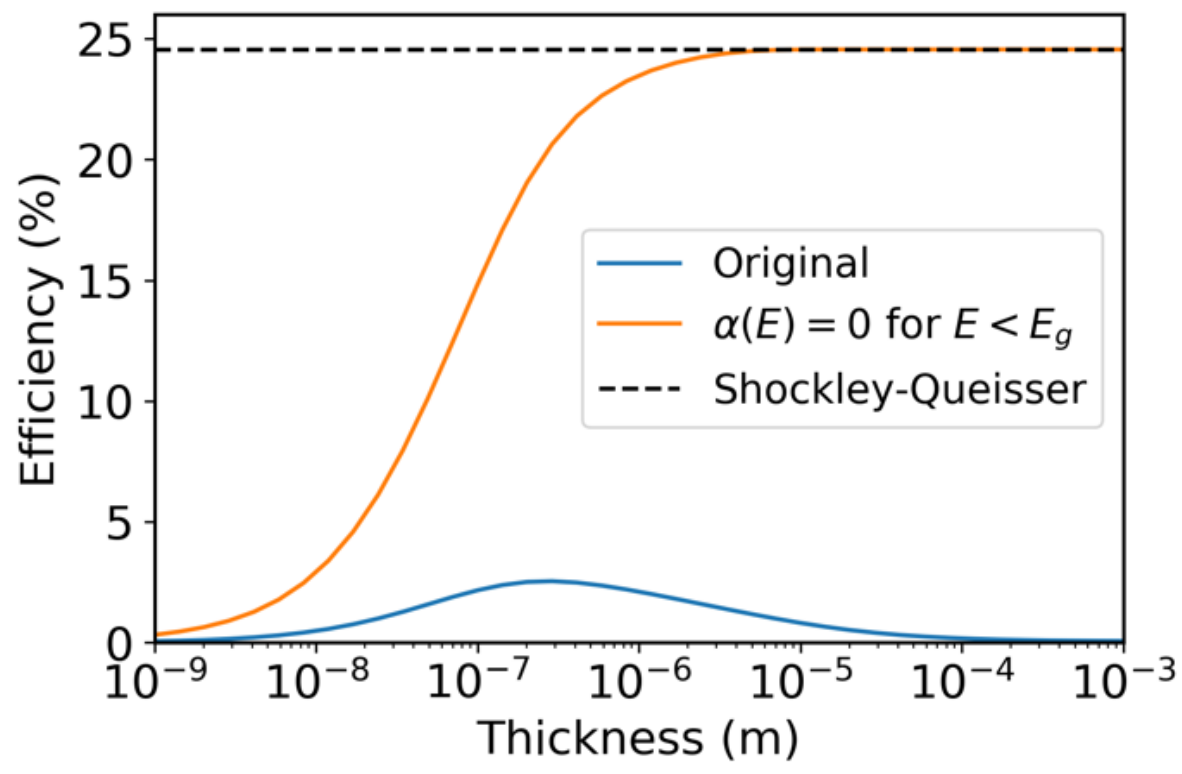

*Figure S2: SLME versus the thickness L, both for the absorption spectrum from original dielectric function, as well as the one where we have set the absorption to zero below the band gap. The Shockley-Queisser limit of the calculated band gap of CuAu-$CuInSe_2$ is plotted as a reference. Note that all calculations were performed for a solar cell at temperature 300K.*